\documentclass[11pt]{article}
\usepackage[latin1]{inputenc}
\usepackage[english]{babel}
\usepackage[namelimits]{amsmath}
\usepackage{authblk}
\usepackage{amssymb}
\usepackage{amsmath}
\usepackage{amsthm}

{\tiny}

\begin{document}
\title{{\bf A proposal for Heisenberg uncertainty principle and STUR for curved backgrounds: an application to white dwarf, neutron stars and black holes.}}
\author[1,2,3]{S. Viaggiu\thanks{s.viaggiu@unimarconi.it and viaggiu@axp.mat.uniroma2.it}}
\affil[1]{Dipartimento di Fisica Nucleare, Subnucleare e delle Radiazioni, Universit\'a degli Studi Guglielmo Marconi, Via Plinio 44, I-00193 Roma, Italy.}
\affil[2]{Dipartimento di Matematica, Universit\`a di Roma ``Tor Vergata'', Via della Ricerca Scientifica, 1, I-00133 Roma, Italy.}
\affil[3]{INFN, Sezione di Napoli, Complesso Universitario di Monte S. Angelo, Via Cintia Edificio 6, 80126 Napoli, Italy.}
\date{\today}\maketitle
\begin{abstract}
\noindent After a critical overview of the Generalized Uncertainty Principle (GUP) applied to compact objects, we 
propose a texture of Heisenberg uncertainty principle in curved spacetimes (CHUP). CHUP allows to write down physically motivated STUR (spacetime uncertainty relations) 
in a generic background for a non commutative 
spacetime in terms of tetrad variables. 
In order to study possible quantum effects for compact astrophysical objects as white dwarf, neutron stars and black holes,
an expression for quantum fluctuations is outlined.
As a result, contrary to GUP-based claims, we found no evidence for quantum effects 
concerning equilibrium equation and critical mass $M_c$ for white dwarf and neutron stars. Conversely, our expression for CHUP confirms
that general relativistic effects strongly reduce the Oppenheimer-Volkoff Newtonian limit for very compact astrophysical objects as neutron stars. In particular, we found that for a degenerate relativistic Fermi gas, the maximum mass decreases for increasing compactness of the star with a minimum critical mass
$M_c\simeq 0.59 M_{\odot}$ at the Buchdahl limit. Finally, we study possible non commutative effects near the event
horizon of a black hole.
\end{abstract}

PACS Number(s): 04.20.Cv, 04.60.-m, 04.40.Dg, 04.70.Dy, 97.60.Jd

\section{Introduction}

The formulation of a well posed quantum gravity theory is one of the biggest problem in modern physics. The main difficulty is due to the fact that, contrary to the
ordinary quantum mechanics, no quantum signature provided by experiments is actually at our disposal. Quantum gravity is expected to emerge at the fundamental
Planck length $L_P$, with 
$L_P=\sqrt{\frac{G\hbar}{c^3}}\simeq 10^{-35}$ meters. The Planck length $L_P$ is well below the scales that can be tested with the actual technology. A possible 
approach to circumvent this issue is to study possible quantum gravity effects when very strong gravitational fields are present. In the universe, astrophysical objects
where the density is so hight that ordinary matter can exist only in a degenerate state, namely degenerate Fermi gas, are represented by white dwarf and neutron stars.
As well known, white dwarf stars represent the final fate of stars with mass $M$ less than the Chandrasekhar \cite{1} one (Chandrasekhar limit), namely $M_{c}$, where
$M_c\simeq 1.44 M_{\odot}$, and with a density $\rho$ of the order of $10^{4}-10^7\;g/cm^3$. A star with  mass $M$ greater than the Chandrasekhar limit 
(dubbed Oppenheimer-Volkoff limit) can reach an equilibrium state
with protons and electrons crushing together and thus producing neutrons. The critical mass for a neutron star is about $\sim 1-3 M_{\odot}$, 
the uncertainty depending on the unknown
equation of state. The typical density of a neutron star is about $10^{14}\;g/cm^3$. Such objects, together with black holes, are thus natural candidates 
to explore possible relativistic and quantum gravity effects on macroscopic scales. In particular, many papers  (see for example
\cite{2,3,4} and references therein) using the well known GUP, firstly proposed in \cite{5,6} and in \cite{7} with a "gedanken experimenten",
are present in literature. In fact, it is possible to 
calculate the Chandrasekhar limit for a white dwarf (see for example \cite{2,3,4}) by using GUP. As a consequence, the authors in \cite{2,3,4} found the unrealistic result
that Chandrasekhar limit disappears after adopting GUP. This fact poses serious doubts on the physical viability of GUP: alternative possibilities must be explored.
The aim of this paper is to propose an heuristic alternative to GUP-based approach and then to apply it to compact astrophysical objects. In section 2 we briefly review the GUP approach and its application to astrophysical objects. In section 3 tetrad formalism is used to write down Heisenberg uncertainty relations in a curved spacetime
(CHUP). In section
4 STUR are written in terms of tetrad formalism and specified to a static spacetime in spherical coordinates. In sections 5,6,7 we apply the machinery of the section above to study relativistic and quantum effects for compact stars, while section 8 is devoted to the study of STUR near the event horizon of a black hole. Finally, section 9 is devoted to some 
conclusions and final remarks.

\section{Minimum length: beyond GUP-based approach}

It is widely accepted that quantum gravity must incorporate a minimum detectable length of the order of the fundamental Planck length $L_P$. The idea of a minimum length 
arising from the localizing procedure of spacetime events can be traced back to Wigner \cite{5}. In particular, Wigner noticed that the resolution of close network of 
points requires a sufficiently large amount of energy. Under the crucial assumption of spherical symmetry
with the uncertainties of the Cartesian coordinates of the same magnitudo, i.e.
\begin{equation}
\Delta x \sim \Delta y \sim \Delta z \sim \Delta r,
\label{1c}
\end{equation}
this idea has been put in an explicit formula by Mead \cite{6}:
\begin{equation}
\Delta r \geq L_P.
\label{2c}
\end{equation}
An equivalent result has been obtained in a string context in \cite{7,8}, in \cite{9} with a gedanken experimenten and in \cite{10} by model independent arguments.
The GUP in the simplest and original form is given by:
\begin{equation}\
\Delta x\Delta p\geq\frac{1}{2}\left[\hbar+\frac{\alpha L_P^2 {\Delta p}^2}{\hbar}\right],
\label{3c}
\end{equation}
where $\alpha$ is an unspecified constant that is expected to be of the order of unity. As evident, GUP (\ref{3c}), as also  noted in \cite{11},
implicitely assumes the spherical symmetry in the localizing procedure. However, the Mead work in \cite{6} clearly shows that a minimum length uncertainty
can be obtained independently from GUP. The GUP (\ref{3c}) has been implemented in terms of modified commutators for the operators $p,x$ in 
\cite{12} by a deformation of the usual Heisenberg algebra. GUP emerges in several attempts to formulate a quantum gravity theory as string and double special 
relativity to cite someone. 
To cite some recent applications, in \cite{r1} has been shown that GUP provides quantum corrections to the Newtonian potential, while in \cite{r2} 
the authors show that GUP is in agreement with the Lorentz-violating extension of the Standard Model. In \cite{r3} GUP is used in the context of the
corpuscolar description of black holes. Finally, in \cite{r4} it is shown that GUP provides a maximal acceleration framework.\\
However, despite the positive role played by GUP in a quantum gravity framework and its interesting features,  
such  modified commutation relations present many unsolved theoretical problems (see for example \cite{13} and references therein for a recent attempt to
alleviate these issues). In fact, GUP implies a violation of the Poincar\'e invariance in the relativistic limit and a breackdown of the equivalence principle 
to cite the most relevant. In particular, a modification of Poincar\'e invariance, well constrained by standard QED physics and 
astrophysical data from gamma-ray-burst and gravitational waves physics, poses serious restrictions to the validity of GUP. It is important to notice that a minimum
uncertainty certainly leads to a minimum length, but, as firstly noticed in \cite{4a} and further in \cite{14}, a minimum uncertainty is not mandatory to 
obtain a minimum length in the operators spectrum. In fact, in the well known Doplicher-Fredenaghen-Roberts (DFR) model \cite{4a}, it is argued that there is not a lower limitation
on the uncertainty of a single coordinate in the localizing procedure. As a consequence, in DFR model the coordinates become self-adjoint operators where the 
Poincar\'e invariance is lifted to a quantum level for the commutators among coordinates itself. A minimum length appears in such a scenario \cite{15} without the necessity of
a minimum uncertainty in a single coordinate. This approach is debated in \cite{15} (see also \cite{16} for a response). The point of debat is that GUP could be only suitable 
for localization of particles in 
first quantization. However, relativistic quantum matter must be depicted in terms of quantum fields.
In relativistic quantum field theory the interest is mainly focused on the localizability of quantum 
field theoretical observables rather than on the limitations on the measurement of the position of a given particle (see for example \cite{6a} for a more complete discussion
on this issue).\\
Another point of debat concerning GUP is that, as shown for example in \cite{2,3,4}, when applied to white dwarf physics, GUP (\ref{3c}) removes
the Chandrasekhar limit, an exotic result in contradiction with all astrophysical data at our disposal. In \cite{2} one realizes that the issue can be solved by taking
$\alpha <0$ with the undesired result that the algebra becomes commutative at Planckian scales, once again an exotic result. In \cite{4} the authors show that 
the introduction of a positive small cosmological constant can solve the issue, i.e. stated in other words the cosmological constant 'protects' the 
Chandrasekhar limit. Also this solution, although intriguing, is rather questionable. In fact, white dwarf physics is not determined by cosmological scales but rather 
by local ones located inside galaxies. In particular, stars evolve inside dense regions where matter is collapsing and as a consequence
local expansion parameter $\theta$ is expected to be negative. Stated in other words, in a dense region where stars form and evolve, the very small positive contribution 
due to the cosmological constant to the local measured expansion parameter is completely negligible and as a result for the local expansion parameter ${\theta}_d$
in such regions we have ${\theta}_d<0$. Hence, we point out that it is the whole negative expansion parameter ${\theta}_d$ that should enter in the modified GUP (EGUP) in
\cite{4} rather than the negligible positive contribution due to the cosmological constant. Another point of discussion, not obviously restricted to GUP
(\ref{3c}), is that the quantity $\Delta x$ is not suitable in a curved context, where distances are measured in terms of the metric tensor $g_{\mu\nu}$. It is thus 
desirable to work with objects having an intrinsic meaning in a curved spacetime, where the metric tensor enters into action.\\
For all resonings above it is thus evident the necessity to explore alternative textures of the Heisenberg uncertainty relations in a curved spacetime. In the next 
section, in order to explore possible relativistic and quantum signatures for the physics of compact objects, we propose a new 
heuristic but physically motivated texture of the Heisenberg uncertainties and we apply them to build physically motivated STUR in a generic background.

\section{Tetrad formalism and Heisenberg uncertainty relations}

Tetrad formalism \cite{2a,3a} represents a powerful tool of general relativity allowing to separate general coordinate transformations 
from trasformations of reference frame. Tetrads represent a local basis for the tangent bundle and more precisely
are coefficients expansion of components of a orthonormal basis
over the differentials expressed in an arbitrary set of assigned coordinates. In the following we use an orthonormal tetrad basis,
representing locally an inertial frame at each point, in such a way that the scalar product of the axes 
${\gamma}_{(a)}$ it gives a Minkowskian metric
${\eta}_{(a)(b)}=diag(-1,1,1,1)$ with ${\gamma}_{(a)}.{\gamma}_{(b)}={\eta}_{(a)(b)}$. For a spacetime equipped with Lorentzian
metric $g_{\mu\nu}$,  $e^{\mu}_{(a)}$ denotes a basis of contravariant vectors, with $\{(a)\}=\{0,1,2,3\}$ tetrad indices, while
$\{\mu\}=\{0,1,2,3\}$ denote tensor indices. We have the standard properties:
\begin{equation}
e_{(a)\mu}=g_{\mu\nu}e^{\nu}_{(a)},\;\;\;\;e^{\mu}_{(a)}e_{(b)\mu}={\eta}_{(a)(b)}.
\label{1}
\end{equation}
In term of the orthonormal tetrad basis (\ref{1}), we can write any vector or tensor field in terms of tetrad components:
$A_{(a)}=e_{(a)\mu}A^{\mu}$, $T_{(a)(b)}=e^{\mu}_{(a)}e^{\nu}_{(b)}T_{\mu\nu}$. Finally, the line element becomes:
\begin{equation}
ds^2={\eta}_{(a)(b)}{\omega}^{(a)}{\omega}^{(b)},\;\;\;{\omega}_{(a)}=e_{(a)\mu}dx^{\mu},
\label{2}
\end{equation}
where $\{x^{\mu}\}$ are the coordinates chosen for $g_{\mu\nu}$ and ${\omega}_{(a)}$ is the projection of the differential in the chosen 
coordinates $\{x^{\mu}\}$ along the tetrad frame. Hence, we can obtain the expression for the local 
tetrad coordinates ${\eta}^{(a)}$, with $d{\eta}^{(a)}=e_{\mu}^{(a)}dx^{\mu}$, by a integration along the tetrad basis directions, i.e.
${\eta}^{(a)}=\int e_{\mu}^{(a)}dx^{\mu}$.\\
In order to obtain STUR in a generic background, we need a physically reasonable expression for the Heisenberg uncertainty
principle and tetrad formalism can help us to obtain such a expression. To start with, we can introduce the directed
derivative ${\partial}_{(a)}$ as
\begin{equation}
{\partial}_{(a)} = {\gamma}_{(a)}.{\bf\partial} = e_{(a)}^{\mu}\frac{\partial}{\partial x^{\mu}} =
e_{(a)}^{\mu}{\partial}_{\mu}.
\label{3}
\end{equation}
Note that the directed derivative ${\partial}_{(a)}$ is defined in terms of tetrad
indices only and thus represents
a tetrad frame four-vector.
We can thus define a tetrad spatial momentum $p^{(A)}$ ($\{A\}=\{1,2,3\}$) as:
\begin{equation}
p_{(A)}=-\imath\hbar {\partial}_{(A)}.
\label{4}
\end{equation}
Since by definition $d{\eta}^{(a)}=e_{\mu}^{(a)}dx^{\mu}$ and using the second equation of
(\ref{1}), we obtain
\begin{eqnarray}
& & p_{(A)}({\eta}_{(A)})=
-\imath\hbar\;e^{\mu}_{(A)}{\partial}_{\mu}\int e_{\nu(A)}dx^{\nu}=
-\imath\hbar {\eta}_{(A)(A)}=-\imath\hbar,\nonumber\\
& & p^{(A)}({\eta}^{(A)})=
-\imath\hbar\;e^{\mu(A)}{\partial}_{\mu}\int e_{\nu}^{(A)}dx^{\nu}=
-\imath\hbar {\eta}^{(A)(A)}=-\imath\hbar, \label{h1} 
\end{eqnarray}
Properties (\ref{h1}) follow because (\ref{3}) represents the directional derivative along the axes ${\gamma}_{(A)}$ and after 
choosing the one represented by the direction pointed by 
${\eta}^{(A)}$. As a result, such a derivative is independent of the choice of coordinates. Stated in other word, (\ref{3}) is the dual version of the 
relation $d{\eta}^{(a)}=e_{\mu}^{(a)}dx^{\mu}$ (co-frame), i.e. we have
\begin{equation}
d{\eta}^{(A)}=e_{\mu}^{(A)}dx^{\mu}\rightarrow\frac{dx^{\mu}}{d{\eta}^{(A)}}=e_{(A)}^{\mu}\rightarrow
{\partial}_{(A)}=\frac{dx^{\mu}}{d{\eta}^{(A)}}{\partial}_{\mu}
\label{h2}
\end{equation}
Consequently, $p^{(A)}$ and ${\eta}^{(A)}$ are conjugate variables.
We are thus legitimated to write down the following Heisenberg uncertainty relations:
\begin{equation}
\Delta p^{(A)}\Delta{\eta}^{(A)}\geq \frac{\hbar}{2}. 
\label{5}
\end{equation}
Commutators for (\ref{5}) can be written as:
\begin{equation}
\left[{\eta}^{(A)},p^{(A)}\right]= \imath\hbar. 
\label{5b}
\end{equation}
After denoting with $E$ the quasi-local energy in a given compact region, we can also write:
\begin{equation}
\Delta E\Delta{\eta}^{(0)}\geq\frac{\hbar}{2}.
\label{6}
\end{equation}
Note that spacetime metric indirectly enters in the proposals (\ref{5}) and (\ref{6}) for CHUP: the metric is explicitely present in the  expressions for ${\eta}^{(A)}$ variables.\\
Once CHUP have been obtained, we can use relations (\ref{5}) and (\ref{6}) to get physically motivated STUR in a generic asymptotically flat spacetime.
Since in the following we are mainly interested to static backgrounds, 
we must obtain the general tetradic relations above in static spacetimes. 
To this purpose, 
we consider the most general expression suitable for a static spacetime in spherical coordinates $\{r,\theta,\phi\}$:
\begin{equation}
ds^2=-f(r)dt^2+h(r)dr^2+r^2\left[d{\theta}^2+\sin^2(\theta)d{\phi}^2\right],
\label{7}
\end{equation}
with $f(r)>0$ and $h(r)>0$ and $r>0, \phi\in[0, 2\pi), \theta\in[0,\pi]$. 
In the vacuum we have the Schwarzschild solution with $f(r)h(r)=1$ and
$f(r)=1-\frac{2GM}{c^2 r}$ with $M$ the ADM mass of the spacetime. The tetrad associated with (\ref{7}) is:
\begin{eqnarray}
& & e_{\mu}^{(0)}=\left(\sqrt{f(r)},\;0,\;\;0,\;\;0\right), \label{8}\\
& & e_{\mu}^{(1)}=\left(0,\;\;\sqrt{h(r)},\;\;0,\;\;0\right), \nonumber\\
& & e_{\mu}^{(2)}=\left(0,\;0,\;\;r,\;\;0\right), \nonumber\\
& & e_{\mu}^{(3)}=\left(0,\;0,\;\;0,\;\;r\sin(\theta)\right). \nonumber
\end{eqnarray}
In order to describe a localizing experiment in such a background, we must obtain the energy and momentum of a localizing photon. To this purpose,
for the metric (\ref{7}) the geodesic equation for a photon is given by $ds^2=0$. For a photon we also have $E^2=c^2 p^2=h^2\nu^2$.
Moreover, thanks to (\ref{7}) in a reference with proper time $T$ for the impulse of the localizing photon we have:
\begin{eqnarray}
& & v^1=\dot{r},\;\;v^2=\dot{\theta},\;\;v^3=\dot{\phi},\;\;\dot{()}=\frac{d}{dT},\:\:dT=\sqrt{f(r)} dt\nonumber\\
& & v_1=h(r)\dot{r},\;\;v_2=r^2\dot{\theta},\;\;v_3=r^2\sin^2(\theta)\dot{\phi},\nonumber\\
& & E=\frac{h\nu}{c^2}\left[v^1 v_1+v^2 v_2+v^3 v_3\right], \label{p} 
\end{eqnarray}
where the time $T$ has an intrinsic meaning. 
In a tetrad frame with
\begin{equation}
ds^2=-{d{\eta}^{(0)}}^2+{d{\eta}^{(1)}}^2+{d{\eta}^{(2)}}^2+{d{\eta}^{(3)}}^2, \label{pp}
\end{equation}
the photon energy is:
\begin{equation}
E=h\nu\left[{\dot{\eta}}^{(1)2}+{\dot{\eta}}^{(2)2}+{\dot{\eta}}^{(3)2}\right],\;
\dot{()}=\frac{d}{\;\;d{\eta}^{(0)}}, \label{k} 
\end{equation}
where (\ref{p}) and (\ref{k}), thanks to (\ref{8}), represent the same expression with $d{\eta}^{(0)}=dT$. 

\section{STUR in tetrad formalism and spherical coordinates}

To start with, we consider a rectangular Planckian localizing box along the tetrad axis ${\gamma}_{(a)}$. Thanks to our definition for
$d{\eta}^{(a)}=e_{\mu}^{(a)}dx^{\mu}$, for the proper area $A$ and the proper
volume $V$ of our Planckian box we can write the following expressions:
\begin{eqnarray}
& & A\sim \Delta{\eta}^{(1)}\Delta{\eta}^{(2)}+\Delta{\eta}^{(1)}\Delta{\eta}^{(3)}+\Delta{\eta}^{(2)}\Delta{\eta}^{(3)},
\label{9}\\
& & V \sim \Delta{\eta}^{(1)}\Delta{\eta}^{(2)}\Delta{\eta}^{(3)}.\label{10}
\end{eqnarray}
Mimicking the procedure in \cite{5a,6a}, a further step is to consider a sufficient condition 
ensuring that a black hole does not form during the experiment. To this purpose we use Penrose's inequality (see \cite{5a,6a}
and references therein): {\it for asymptotically flat data, horizons form if and only if}:
\begin{equation}
A < 16\pi\frac{G^2}{c^4}m^2,
\label{11}
\end{equation}
where $m$ is the mass inside $A$. Thanks to (\ref{11}) we can write a sufficient condition for no black hole 
formation during the localizing experiment:
\begin{equation}
A\geq 16\pi \frac{G^2}{c^4}m^2.
\label{12}
\end{equation}
After denoting with $\Delta E=mc^2$ the amount of quasi-local energy within $A$, for a single particle with momentum
$p^{(A)}$, that in the present case is the localizing photon with energy given by (\ref{p}) (or (\ref{k}))
the following inequality holds:
\begin{equation}
{\Delta E}^2\geq \frac{c^2}{3}\left[{(\Delta p^{(1)})}^2+
{(\Delta p^{(2)})}^2+{(\Delta p^{(3)})}^2\right].
\label{13}
\end{equation}
By using the inequalities (\ref{5}), (\ref{6}), (\ref{12}), (\ref{13}), for the STUR we obtain (see \cite{5a,6a}):
\begin{eqnarray}
& & {(c\Delta_\omega{\eta}^{(0)})}^2\left(\Delta_\omega{\eta}^{(1)}\Delta_\omega {\eta}^{(2)}+
\Delta{\eta}^{(1)}_\omega \Delta_\omega{\eta}^{(3)}+
\Delta_\omega {\eta}^{(2)}\Delta_\omega {\eta}^{(3)}\right) \geq
L_{P}^4, \nonumber\\
& & {(\Delta_\omega {\eta}^{(1)})}^2 {(\Delta_\omega {\eta}^{(2)} )}^2 {(\Delta_\omega {\eta}^{(3)} )}^2 \geq\nonumber\\ 
& & \geq L_{P}^4
\left(\Delta_\omega {\eta}^{(1)}\Delta_\omega {\eta}^{(2)}+
\Delta_\omega{\eta}^{(1)}\Delta_\omega{\eta}^{(3)}+\Delta_\omega {\eta}^{(2)}\Delta_\omega{\eta}^{(3)}\right) \label{15}.
\end{eqnarray}
Since the inequalities above have a quantum interpretation,
we added the subscript $\omega$ to emphasise their dependence on quantum states. 
By adopting the following inequalities:
\begin{eqnarray}
& &(a+b+c)^2\geq ab+bc+ac, \label{algebraic}\\
& &(ab+bc+ac)^3\geq a^2 b^2 c^2,\nonumber
\end{eqnarray}
the (\ref{15}) become:
\begin{eqnarray}
& & c\Delta_\omega {\eta}^{(0)}\left[\Delta_\omega {\eta}^{(1)}+\Delta_\omega {\eta}^{(2)}
+\Delta_\omega {\eta}^{(3)}\right]\geq L_P^2,\label{16}\\
& & \Delta_\omega {\eta}^{(1)}\Delta_\omega {\eta}^{(2)}+\Delta_\omega {\eta}^{(1)}\Delta_\omega {\eta}^{(3)}+
\Delta_\omega {\eta}^{(2)}\Delta_\omega {\eta}^{(3)}\geq L_P^2. \label{17}
\end{eqnarray}
The STUR (\ref{16})-(\ref{17}) are the ones present in \cite{4a} but expressed in terms of the variables ${\eta}^{(a)}$. It is important to note that (\ref{15})
and (\ref{16})-(\ref{17}) are valid in any asymptotically flat spacetime.
At this level of treatment, for applications to static astrophysical objects,
we can write down physically motivated uncertainties relations adapted to spherical coordinates.
The commutators implying the STUR  (\ref{16})-(\ref{17}) are similar to the ones in \cite{4a} but expressed in terms of
${\eta}^{(a)}$ variables. Once these are obtained, we need to express commutators in spherical coordinates, the coordinates adapted to the Killing vectors of 
a spherically symmetrical spacetime.
In order to express 
(\ref{17}) in spherical coordinates, we use the following reasonable approximations 
for our localizing experiment:
\begin{eqnarray}
& & \Delta {\eta}^{(0)} \sim e_0^{(0)}\Delta x^0=c\sqrt{f(r)}\Delta t,\label{18}\\
& & \Delta {\eta}^{(1)} \sim e_r^{(1)}\Delta r=\sqrt{h(r)}\Delta r, \label{19}\\
& & \Delta {\eta}^{(2)} \sim e_{\theta}^{(2)}\Delta\theta=r\Delta\theta, \label{20}\\
& & \Delta {\eta}^{(3)} \sim e_{\phi}^{(3)}\Delta\phi=r|\sin(\theta)|\Delta\phi. \label{21}
\end{eqnarray}
With (\ref{18})-(\ref{21}), equations (\ref{16})-(\ref{17}) become:
\begin{eqnarray}
& & c\sqrt{f(r)}\Delta t\left[\sqrt{h(r)}\Delta r+r\Delta\theta+r|\sin(\theta)|\Delta\phi\right]\geq  L_P^2,\label{22}\\
& & r\left[
\sqrt{h(r)}\Delta r\Delta\theta+|\sin(\theta)|\sqrt{h(r)}\Delta r\Delta\phi+r|\sin(\theta)|\Delta\theta\Delta\phi\right]
\geq  L_P^2.\label{23}
\end{eqnarray}
For a better understanding of the approximations made, note that in practice we have approximated the area of the localizing region with
the one of the spherical region enclosing the box itself. In fact we have
\begin{eqnarray}
& &\Delta{\eta}^{(1)}\Delta{\eta}^{(2)}+\Delta{\eta}^{(1)}\Delta{\eta}^{(3)}+
\Delta{\eta}^{(2)}\Delta{\eta}^{(3)}\sim \label{24}\\
& &\sim \Delta A_{r-\theta}+\Delta A_{r-\phi} +\Delta A_{\phi-\theta},\nonumber
\end{eqnarray}
where $\Delta A_{i-j}, \{i,j\}=\{r,\theta,\phi\}$ is the area of the 'face' $i-j$.
With the metric (\ref{7}), we have $\Delta A_{i-j}\sim\sqrt{g_{ii}}\sqrt{g_{jj}}\Delta x^i\Delta x^j$ and the (\ref{22})-(\ref{23})
emerge.\\
Concerning the term $\sin\theta$, consider the localizing experiment with a spherical box placed 
symmetrically with respect to the
equatorial plane.
To obtain $A_{\phi-\theta}$ we must perform the following integral:
\begin{equation}
A_{\phi-\theta}=r^2\int_{-\Delta\phi}^{\Delta\phi}d\phi
\int_{\frac{\pi}{2}-\Delta\theta}^{\frac{\pi}{2}+\Delta\theta}|\sin\theta| d\theta.
\label{a1}
\end{equation}
We obtain:
\begin{eqnarray}
& & A_{\phi-\theta}=r^2\;2\Delta\phi\left[
\cos\left(\frac{\pi}{2}-\Delta\theta\right)-\cos\left(\frac{\pi}{2}+\Delta\theta\right)\right]=\nonumber\\
& & -4r^2\Delta\phi\sin\left(\frac{\pi}{2}\right)\sin(-\Delta\theta)\sim 4r^2\Delta\phi\Delta\theta.
\label{a2}
\end{eqnarray}
A similar expression arises for $A_{r-\theta}$ where an integration over $\theta$ is also necessary.
This 'privileged' spherical box can be obviously transported, for example, at the north pole o rotated along the $\phi$ axis by $\phi$ and still remains of the same shape and 
proper dimension. As a consequence, without loss of generality, we can set
$|\sin(\theta)|=1$ and hence the (\ref{22})-(\ref{23}) reduce to
\begin{eqnarray}
& & c\sqrt{f(r)}\Delta t\left[\sqrt{h(r)}\Delta r+r\Delta\theta+r\Delta\phi\right]\geq  L_P^2,\label{25}\\
& & r\left[
\sqrt{h(r)}\Delta r\Delta\theta+\sqrt{h(r)}\Delta r\Delta\phi+r\Delta\theta\Delta\phi\right]
\geq  L_P^2.\label{26}
\end{eqnarray}
The next step is to give an operatorial meaning \cite{7a} to the STUR (\ref{25})-(\ref{26}). The problem, as well known, is to give an operatorial meaning to angular spherical
coordinates $\theta,\phi$. To this purpose, we limit to observe (see for example \cite{8a,9a} and references therein) that it is always possible to found representations where
STUR in spherical coordinates make sense. Fortunately, for the purpose of this paper, we only need physically motivated formulas for Heisenberg uncertainty relations for 
the  momentum and quantum fluctuations in a spherical background. Concerning the momentum, in order to study the stellar
equilibrium configuration, we are interested in a crude but reasonable approximate formula for the radial momentum for a degenerate Fermi gas capable to struggle with 
pressure of gravity. Thanks to (\ref{5})
and (\ref{19}) we obtain:
\begin{equation}
\Delta P^{(r)}\sim \frac{\hbar}{2\sqrt{h(R)}\Delta r}. 
\label{27}
\end{equation}
We need to evaluate formula (\ref{27}) at the boundary of a star where $f(R)=\frac{1}{h(R)}=1-\frac{2GM}{c^2 R}$. Moreover, at $r=R$ we can write $\Delta r\sim R$. Consequently
we obtain:
\begin{equation}
\Delta P^{(r)}\sim \frac{\hbar}{2R}\sqrt{1-\frac{2GM}{c^2 R}}.
\label{28}
\end{equation}
Concerning quantum fluctuations, from (\ref{6}) and (\ref{18}) evaluated at $r=R$ we can write
\begin{equation}
\Delta E\sim\frac{\hbar}{2\Delta\eta^{(0)}}\sim \frac{\hbar}{2\Delta t\sqrt{1-\frac{2GM}{c^2 R}}}.
\label{29}
\end{equation}
To evaluate $\Delta t$, consider STUR (\ref{25}) in a spherical localizing state \cite{4a} where all uncertainties have the same magnitudo. In particular, from (\ref{25})
evaluated at $r=R$ we deduce that\footnote{This is because in (\ref{25}) we have $f(R)h(R)=1$ and the other uncertainties in (\ref{25}) in a spherical 
localizing state have the same magnitudo of the first term $\Delta r\;c\Delta t$.} $c\Delta t\sim \Delta R\sim R$ 
and as a result for quantum fluctuations we obtain the estimation 
\begin{equation}
\Delta E=\frac{c^4\chi L_P^2}{2GR\sqrt{1-\frac{2GM}{c^2 R}}},
\label{30}
\end{equation} 
where the constant $\chi$ depends on the STUR (\ref{25}) and (\ref{26}) and it is expected of the order of unity. As a first consideration for (\ref{30}), note that the 
chosen coordinates
are valid for $r\geq R_s$, where $R_s$ denotes the Schwarzschild radius. We could also think to extend formula (\ref{30}) inside the event horizon of a black hole, as 
usual, by taking $\sqrt{1-\frac{2GM}{c^2 R}}\rightarrow\sqrt{\left|1-\frac{2GM}{c^2 R}\right|}$ and thus fluctuations remain real inside a black hole.
In practice, quantum fluctuations are very strong at Planckian
scales, but them are watered down on bigger and bigger scales, represented by the factor $R$ at the denominator, corrected by the relativistic compactness factor. 
Similar estimations for quantum fluctuations in a 
cosmological context have been obtained in \cite{10a,11a,12a}. A further feature of expression (\ref{30}) is that it is diverging at the event horizon $R_s$ of a black hole. 
This phenomenon is a consequence of redshift effects caused by the infinite redshift surface at $R_s$. Thus graviton modes
misured by a distant observer, when traced back to the horizon and due to frozen time $t$ at $R_s$,
all have divergent frequency. This effect is very similar to the well known trans Planckian problem for
Hawking radiation and it is thus expected for near-horizon quantum fluctuations.\\
With estimations (\ref{28}) and (\ref{30}) we are ready to capture possible relativistic and 
quantum effects for compact astrophysical objects 

\section{Chandrasekhar limit from Heisenberg uncertainty relations}

White dwarf stars are made of electrons in a degenerate state depicted in terms of a Fermi gas at zero temperature. The pressure of degenerate electrons supports the equilibrium of the star versus the pressure due to the attractive gravity. The degenerate Fermi gas is often derived, in an heuristic way, in terms of the usual Heisenberg uncertainty principle.
This can be found, for example, in \cite{2,4} in terms of GUP. Since we are interested in the study 
of our proposed CHUP in relation to GUP (EGUP or similar), we repeat this derivation. First of all, as usual, we consider the star composed of 
one-half helium atoms with one proton and one neutron and $N$ degenerate electrons of mass $m_e$. Since for neutron mass $m_n$ and proton mass $m_p$ we have
$m_n\simeq m_p$ with $m_n>>m_e$, by denoting with $M$ the total mass of the star we have
\begin{equation}
n=\frac{N}{V}=\frac{M}{2m_n V},\;\;V=\frac{4\pi}{3}R^3.
\label{4c}
\end{equation}
A degenerate Fermi gas is specified by taking 
\begin{equation}
\Delta x=\alpha_e^{-1} n^{-\frac{1}{3}},
\label{5c}
\end{equation}
where $\alpha_e$ is a positive constant that must be fixed in order to obtain the Newtonian value for the critical mass $M_c$. The Fermi momentum is obtained by means of
the Heisenberg relation $\Delta p=p_F=\hbar/(2\Delta x)$, with $\Delta x$ given by (\ref{5c}). Hence, for the non-relativistic energy $E_F$ of $N$ degenerate electrons we have, by using
(\ref{4c}) and (\ref{5c}):
\begin{equation}
E_F=\frac{N p_F^2}{2m_e}=
\frac{\alpha_e^2\hbar^2}{16 m_e m_n^{\frac{5}{3}}{\left(\frac{8\pi}{3}\right)}^{\frac{2}{3}}}\frac{M^{\frac{5}{3}}}{R^2}.
\label{6c}
\end{equation}
For the binding energy $E_g$ we take the usual expression $E_g=-GM^2/R$ and for the total energy $E_T$ 
we have $E_T=E_F+E_g$. The minimum of $E_T$ with respect to $R$ is obtained
for
\begin{equation}
R=\frac{\alpha_e^2\hbar^2{\left(\frac{3}{8\pi}\right)}^{\frac{2}{3}}}{8m_e m_n^{\frac{5}{3}}G M^{\frac{1}{3}}}.
\label{7c}
\end{equation}
The critical mass is obtained when electrons become relativistic. This happens for
\begin{equation}
p_F=\frac{\alpha_e\hbar\; 3^{\frac{1}{3}}M^{\frac{1}{3}}}{2{\left(8\pi\right)}^{\frac{1}{3}}m_n^{\frac{1}{3}}R}
=m_ec.
\label{8c}
\end{equation}
 After solving (\ref{8c}) for $R$ and equating to (\ref{7c}) we obtain
 \begin{equation}
 M_c=\frac{\alpha_e^{\frac{3}{2}} c^{\frac{3}{2}}\hbar^{\frac{3}{2}}}{8G^{\frac{3}{2}}m_n^2}\sqrt{\frac{3}{8\pi}}.
 \label{9c}
 \end{equation}
 The constant $\alpha_e$ can be fixed by requiring that $M_c$ is given by the well known value $M_c=1.44 M_{\odot}$. From (\ref{9c}) we obtain
 $\alpha_e\simeq 6.86728$. A similar calculation has been performed in \cite{2,4} with GUP. In the next sections we outline the derivaton above
 by using (\ref{28}) and (\ref{30}).

\section{General Relativistic case: an heuristic approach}

In the usual heuristic approach to the Chandrasekhar limit and also for the heuristic one of section above with Heisenberg uncertainty relations, the degenerate gas is supposed ideal 
without gravitational interactions. It is thus expected that gravity plays a role in modifying this idealized picture (See for example \cite{17} for a Newtonian treatment of 
gravity effects by using ordinary quantum mechanics formalism and \cite{18} for a general relativistic treatment).
A general relativistic treatment of stellar equilibrium for spherical stars requires a study of Tolman-Oppenheimer-Volkoff (TOV) equation for the hydrostatic equilibrium. Typically,
once a given equation of state is chosen, a numerical integration of TOV is required. Our study in the following
is not focused on exact solutions of TOV equation, but rather
in the use of the approximate formlas (\ref{28}) and (\ref{30}) to study possible 
relativistic-quantum effects induced by a strong gravitational field in a non numerical heuristic way, this representing a test for our proposed
CHUP (\ref{5}) and (\ref{6}). Concerning the approximations made to obtain (\ref{28}) and (\ref{30}), 
they are based on the reasonable estimations (\ref{18})-(\ref{21}). Hence, we expect an approximation error in the numerical estimation of relativistic and quantum corrections 
to the classical picture (\ref{7c}) and (\ref{9c}). It is also expected that for more compact objects as neutron stars this error becomes smaller. In fact, as an example, for very compact stars as the core of neutron stars, the density $\rho\sim 1/r^2$ can be used. Remember that for a static metric (\ref{7}), the total mass $m(r)$ inside a radial radius $r$ is given by the "Newtonian" expression 
\begin{equation}
m(r)=4\pi\int_0^r \rho(u) u^2 du,
\label{10c}
\end{equation}
with $m(R)=M$. For a power law density profile with $\rho\sim 1/r^n$, with $n$ an integer, (\ref{10c}) is convergent for $n\leq 2$. Hence $n=2$ is the maximum allowed value for 
power law density assuring the convergence of (\ref{10c}). With (\ref{7}) we have $h(r)={\left(1-2Gm(r)/(c^2r)\right)}^{-1}$:
it is a matter of fact that for $n=2$ we have $m(r)/r=const.=M/R$ and the approximation (\ref{28}) becomes an exact formula. 
However, as shown in the next calculations, approximations made are enough to capture relativistic and quantum effects.\\
To start with, we repeat exactly the calculations of section above in the non relativistic case. In the present case, with the help of (\ref{28})
for $p_F$ we obtain
\begin{equation} 
p_F=\frac{\alpha_e\hbar n^{\frac{1}{3}}}{2\sqrt{h(R)}},
\label{11c}
\end{equation}
with formula (\ref{4c}) still valid. For the total non relativistic Fermi energy $E_F$ we obtain:
\begin{equation}
E_F=\frac{\alpha_e^2\hbar^2}{16 m_e m_n^{\frac{5}{3}}{\left(\frac{8\pi}{3}\right)}^{\frac{2}{3}}}\frac{M^{\frac{5}{3}}}{R^2}
\left(1-\frac{2GM}{c^2 R}\right).
\label{12c}
\end{equation}
With the expression $E_g=-GM^2/R$ and posing
\begin{equation}
Y_W=\frac{\alpha_e^2\hbar^2}{16 m_e m_n^{\frac{5}{3}}{\left(\frac{8\pi}{3}\right)}^{\frac{2}{3}}},
\label{13c}
\end{equation} 
for the equilibrium relations, instead of (\ref{7c}) we obtain:
\begin{equation}
R^2-\frac{2Y_W R}{GM^{\frac{1}{3}}}+\frac{6Y_W}{c^2}M^{\frac{2}{3}}=0,
\label{14c}
\end{equation}
and after taking the positive root we get the solution
\begin{equation}
R=\frac{Y_W}{GM^{\frac{1}{3}}}\left(1+\sqrt{1-\frac{6G^2 M^{\frac{4}{3}}}{Y_W c^2}}\right).
\label{15c}
\end{equation}
Formula (\ref{15c}) obviously reduces to (\ref{7c}) in the non relativistic case. To evaluate general relativistic effects for white dwarf, note that the square root term in
(\ref{16c}) is $1-\epsilon$ with $\epsilon\sim 10^{-3}$ and can be Taylor expanded. The result is:
\begin{equation}
R=\frac{2Y_W}{GM^{\frac{1}{3}}}-\frac{3G}{c^2}M+0(1).
\label{16c}
\end{equation}
The first term in (\ref{16c}) represents the expression (\ref{7c}), while the second one furnishes the general relativistic correction of the order of approximatively $5 Km$
for $M\simeq M_{\odot}$. This correction is very small and of the order of $0.1\%$.\\ 
Concerning $M_c$, we must pose, according to (\ref{8c}):
\begin{equation}
p_F=\frac{\alpha_e\hbar 3^{\frac{1}{3}}M^{\frac{1}{3}}}{2{\left(8\pi\right)}^{\frac{1}{3}}m_n^{\frac{1}{3}}R}\sqrt{1-\frac{2GM}{c^2 R}}=m_e c.
\label{17c}
\end{equation}
After substituing the (\ref{15c}) in (\ref{17c}) we obtain a unique solution for $M_c$ given by $M_c\simeq 1.438 M_{\odot}$ for a white dwarf composed 
of one-half helium atoms with degenerate electrons. This correction could be confronted with the one obtained in \cite{18} for white dwarf made of
helium atoms by numerically integrating the TOV equation where $M_c\simeq 1.417$. Considering that general relativistic effects are stronger
for heavier atoms composing the white dwarf, our result differs percentage approximatively by a factor $10$. Since our derivation is heuristic and based on
(\ref{28}) and (\ref{30}), this error is acceptable. The approximation could be amended by taking for the surface gravity $g_s$ the relativistic expression
$g_s=GM/(R^2\sqrt{1-2GM(c^2R)})$ instead of the Newtonian one $g=GM/R^2$. In any case, our model captures the fact that general relativistic effects
lower the Newtonian value for $R$ in (\ref{14c}) and $M_c$. This is due to formulas (\ref{11c}) and (\ref{12c}). In fact, it is generally expected that gravity 
struggles with Fermi degenerate pressure. In particular, gravity is attractive for ordinary matter in general relativity, while 
Fermi pressure opposes to gravitational collapse and as a consequence it is right that gravity modifies the Fermi degenerate pressure of the ideal case
where electrons are treated as free non interacting Fermions. This is a known general relativistic effect.\\
This phenomenon is increasing for increasing compactness term $\sqrt{1-\frac{2GM}{c^2 R}}$. It is thus interesting to explore the calculation above in the limit for very compact
objects where $R\rightarrow R_s=2GM/c^2$, where $R_s$ is the Schwarzschild radius. To this purpose, we could apply the machinery above to neutron stars. However, note that 
neutron stars can be endowed with a non vanishing angular momentum. Moreover, the equation of state for a neutron stars is not yet full understood. Nevertheless, 
in first approximation, it is reasonable to assume a neutron star made of $N$ degenerate neutrons and thus neglect angular momentum.\\
To start with, we consider the non relativistic case. The only changes to made are that neutrons rather than electrons are degenerate and that we have (\ref{4c}) but with
$n=\frac{N}{V}=\frac{M}{m_n V}$. Hence, instead of (\ref{12c}) we have 
\begin{equation}
E_F=\frac{\alpha_N^2\hbar^2}{8 m_n^{\frac{8}{3}}{\left(\frac{4\pi}{3}\right)}^{\frac{2}{3}}}\frac{M^{\frac{5}{3}}}{R^2}
\left(1-\frac{2GM}{c^2 R}\right).
\label{18c}
\end{equation}
In (\ref{18c}) we have introduced the normalization constant $\alpha_N$ for neutrons. By taking, for example, a fiducial value for the critical mass of neutrons stars
without the term $\left(1-\frac{2GM}{c^2 R}\right)$, denoted with $M_c^0$,
given by $M_c^0=3 M_{\odot}$, we have $\alpha_N\simeq 4.5$, while for $M_c^0=M_{\odot}$ one obtains $\alpha_N\simeq 2.147$. The precise value 
of $\alpha_N$ plays no role in this paper. After mimicking the non relativistic calculations above, we obtain again formula (\ref{15c}) for the equilibrium 
radius, but with
$Y_W\rightarrow Y_N$ where
\begin{equation}
Y_N=\frac{\alpha_N^2\hbar^2}{8m_n^{\frac{8}{3}}{\left(\frac{4\pi}{3}\right)}^{\frac{2}{3}}}.
\label{19c}
\end{equation}
The important difference with respect to (\ref{15c}) is that existence condition for (\ref{15c}) with (\ref{19c}) is
\begin{equation}
M\leq \frac{c^{\frac{3}{2}}\alpha_N^{\frac{3}{2}}\hbar^{\frac{3}{2}}}{48^{\frac{3}{4}}m_n^2}{\left(\frac{3}{4\pi}\right)}^{\frac{1}{2}}\simeq 0.158 M_c^{0},
\label{20c}
\end{equation}
where $M_c^{0}$ denotes the fiducial critical mass for the case without the term  $\left(1-\frac{2GM}{c^2 R}\right)$ (As stated above, for example with 
$\alpha_N\simeq 4.5$ we obtain $M_c\simeq 3M_{\odot}$). This result shows the strong effects due to general relativity and the facts that equilibrium configurations 
can exist in the non relativistic case only for masses fulfilling the inequality (\ref{20c}). As a consequence, static degenerate neutron stars 
with masses greater than $\simeq 0.158 M_c^0$ can be
only supported by relativistic neutrons. In this case, for the energy $E_F$ of such a degenerate gas we have $E=cN p_F$ where
\begin{equation}
p_F=\frac{\alpha_N\hbar\;3^{\frac{1}{3}}M^{\frac{1}{3}}}{2{\left(4\pi\right)}^{\frac{1}{3}}m_n^{\frac{1}{3}}R}\sqrt{1-\frac{2GM}{c^2 R}}. 
\label{21c}
\end{equation}
For $E_T$ we thus obtain $E_T=E_F-GM^2/R$. 
In this case, the star is in a equilibrium configuration as far as $E_T\geq 0$: for $E_T<0$ the term 
$E_g=-GM^2/R$ dominates and equilibrium disappears. The critical value $M_c$ for the mass is thus obtained for $E_T=0$ thanks to the following implicit equation: 
\begin{equation} 
M_c=\frac{\alpha_N^{\frac{3}{2}} c^{\frac{3}{2}}\hbar^{\frac{3}{2}}}{2^{\frac{3}{2}}G^{\frac{3}{2}}m_n^2}\sqrt{\frac{3}{4\pi}}
{\left(1-\frac{2GM_c}{c^2R}\right)}^{\frac{3}{4}}.
\label{22c}
\end{equation}
It is interesting to note that critical mass $M_c$ for relativistic very compact stars depends on the compactness factor $2GM/(c^2R)$. Hence, for a fixed radius 
$R>R_s$, we have that $M\leq M_c$.
We can thus  easily study the 
(\ref{22c}) in the following way. We can pose $R_s/R=2GM_c/(c^2R)=1/g$. Moreover, we fix a numerical value for $g$ and thus we solve (\ref{22c}). Finally, with 
the so obtained value for $M_c$ we can obtain the equilibrium radius $R=gR_s$. For some numerical example: $g=10,M_c\simeq2.815 M_{\odot}$;
$g=1.5,M_c\simeq 1.35 M_{\odot}$; $g=1.1,M_c\simeq 0.5 M_{\odot}$; $g=1.01,M_c\simeq 0.088 M_{\odot}$. For the Buchdahl limit $R=9/8 R_s$ representing
the minimum equilibrium radius of a star
composed with a perfect fluid with positive non increasing energy density, we obtain the minimum critical mass $M_c\simeq 0.59 M_{\odot}$. It is worth to be noted that formula  (\ref{22c}) can be applied also to degenerate matter different from neutrons, with the gas composed of a certain particle of mass $m_x$. The physical interesting consequence of (\ref{21c}) and 
(\ref{22c}) is that general relativity reduces the pressure of the degenerate gas and in the limit $R\rightarrow R_s$ the gravitational field is so strong that $p_F\rightarrow 0$, i.e.
no degenerate gas can exist in that limit. As a consequence, 
$M_c$ depends on $g$ in such a way that when we are approaching $R_s$, we have that $M_c\rightarrow 0$ . Stated in other 
words, sufficiently near $R=R_s$ only microscopic configurations can exist, provided that matter is in a degenerate state. However, it should be noticed that for $R$ less than 
Buchdahl limit the configuration becomes unstable and, unless unknown states of matter come into action\footnote{Stars with 
$R\in(R_s, 9/8 R_s)$ could exist in some exotic but physically reasonable form as gravastars or quark stars}, a black hole does form.

\section{Relativistic case with quantum fluctuations}

Quantum fluctuations are dictated by term (\ref{30}). For $E_T$ we have:
\begin{equation}
E_T=E_F-\frac{G M^2}{R}+\frac{c^4\chi L_P^2}{2GR\sqrt{1-\frac{2GM}{c^2 R}}}.
\label{23c}
\end{equation}
Independently on the fact that for $E_F$ we can adopt the relativistic or the non relativistic expression, the role of term (\ref{30}) in (\ref{23c})
crucially depends on its magnitudo with respect the term $E_g$. Quantum fluctuations become relevant when their magnitudo is comparable with $E_g$, i.e. for
\begin{equation}
M^2\sim \frac{\chi}{2}\frac{M_p^2}{\sqrt{1-\frac{2GM}{c^2 R}}},
\label{24c}
\end{equation}
where $M_p$ denotes Planck mass. First of all we consider the white dwarf case. In such a case note that $\frac{2GM}{c^2 R}\sim 10^{-3}$. As a consequence, after denoting
with $E_q$ the quantum fluctuations term (\ref{30}), 
it is easy to see 
that for a white dwarf tipically we have $|E_g|/E_q\sim 10^{76}$. Hence, it is evident that for white dwarf, contrary to GUP-based claims, quantum fluctuations
play no role in the determination of the equilibrium radius (\ref{15c}) an critical mass $M_c$.\\
The situation, thanks to the denominator of $E_q$, could be more interesting in the case of more compact objects as neutron stars. Condition (\ref{24c}) can be fulfilled 
provided that 
\begin{equation}
1-\frac{2GM}{c^2 R}\sim \frac{\chi^2}{4}\frac{M_p^4}{M^4}\simeq 10^{-152}.
\label{25c}
\end{equation}
Condition (\ref{25c}) does imply that, after posing $R=R_s+\epsilon$,  $\epsilon\simeq R_s 10^{-152}<<L_P$. Since for a non commutative spacetime with (\ref{25}) and (\ref{26})
in maximal localizing spherical states we have $\Delta R\sim c\Delta t\sim L_P$, there is no hope to observe a macroscopic astrophysical object 
where Planckian fluctuations play a role in 
the equilibrium configuration. Also if we are wiling to accept the possibility to measure a length of the order of $\epsilon$, we should consider states where
$\Delta t\sim L_P^2/(c\epsilon)\sim 10^{70} sec.$ and thus perform experiments during a time much greater than the age of the universe.\\
As a further consideration, we can calculate the maximum mass $M$ satisfying the (\ref{24c}) together with the constraint $\epsilon\geq L_P$ dictated by 
non commutativity. To this purpose, condition (\ref{25c}) can be written as
\begin{equation}
\epsilon=\frac{R_s\frac{\chi^2}{4}\frac{M_p^4}{M^4}}{\left[1-\frac{\chi^2}{4}\frac{M_p^4}{M^4}\right]}.
\label{26c}
\end{equation}
After setting $\epsilon=L_P$ for $M=M_p$ we obtain $\chi=\sqrt{2}$. It is easy to see that condition $\epsilon\geq L_P$ 
is verified only for\footnote{Remember that for $M=M_p$ we have $R_s=L_P$} $M=M_p$. For any mass 
with $M>M_P$ we have $\epsilon<L_P$. Formula (\ref{26c}) clearly shows that only with hypothetical laboratory experiments 
with self-gravitating systems with mass of the order of the Planck one it is possible to explore possible detectable quantum gravity effects concerning the equilibrium configuration.

\section{The black hole case}

Black holes are important astrophysical objects both from experimental and theoretical point of view. Theoretically, black holes provide
a perfect arena to test gravity in the
very strong regime. In particular, thanks to (\ref{25}) and (\ref{26}), we can explore quantum gravity effects near the black hole event horizon. To do this,
we can study  (\ref{25})-(\ref{26}) in the limit for $r\rightarrow R_s$. First of all, we can write down non commutativity for $r>R_s$ where $f(r)h(r)=1$
with $f(r)=1-2GM/(c^2r)$:
\begin{eqnarray}
& & c\Delta t\Delta r+
c\Delta t\sqrt{f(r)}\left[r\Delta\theta+r\Delta\phi\right]\geq  L_P^2,\label{27c}\\
& & r\left[
\sqrt{h(r)}\Delta r\Delta\theta+\sqrt{h(r)}\Delta r\Delta\phi+r\Delta\theta\Delta\phi\right]
\geq  L_P^2.\label{28c}
\end{eqnarray}  
As a first fact note that metric does appear in (\ref{27c}) only in the angular part of STUR. Moreover, note that by approaching $R_s$ we have that 
$f(r)\rightarrow 0$ and $1/h(r)\rightarrow 0$. Hence, at $r=R_s$, expressions (\ref{27c}) and (\ref{28c}) become
\begin{eqnarray}
& &c\Delta t\Delta r\geq L_P^2, \label{29c}\\
& &\Delta r\left[\Delta\theta+\Delta\phi\right]\geq 0.\label{30c}
\end{eqnarray}
Concerning (\ref{29c}), it is the only STUR in spherical coordinates and involving coordinates $t,r$ surviving in that limit. Hence we have:
\begin{equation}
{\left[ct, r\right]}_{R_S}=\imath L_P^2.
\label{31c}
\end{equation}
Another interesting fact is that at the event horizon the surviving STUR (\ref{29c}) is in some sense Minkowskian, i.e. it is not depending on the metric functions.
The (\ref{30c}) clearly shows that spatial part of STUR becomes weaker when approaching the event horizon and vanishes exactly at $r=R_s$. Consequently, 
tha spatial part of STUR leads to a commutative algebra on the event horizon. As an important consequence, a minimum uncertainty for $\Delta r$ 
disappears at $r=R_s$. It is thus possible to conceive an experiment at the event horizon where $\Delta r\rightarrow 0$, provided that, thanks to
(\ref{29c}), $\Delta t\rightarrow\infty$ as measured by an observer at spatial infinity. This new phenomenon is thus peculiar of non commutative geometry at event
horizon when expressed in spherical coordinates. Moreover, a link between this phenomenon and the well known trans Planckian problem for Hawking radiation 
could be established.
It is also interesting to note that, if we adopt STUR (\ref{16})-(\ref{17}) expressed in tetrad variables, physically we have the same situations of the Minkowskian case,
but now expressed in terms of proper variables. All considerations above can be of interest in a quantum gravity context and can be certainly matter for further 
investigations.

\section{Conclusions and final remarks}

This paper is an attempt to write down physically motivated expressions for Heisenberg uncertainty relations in a curved spacetime (CHUP) and thus to test it
to study general relativistic and quantum effects for astrophysical objects beyond the usual GUP-motivated approaches. In fact, as shown
in section 2, GUP furnishes unphysical results when applied, for example, to white dwarf physics. This is mainly due to the fact that GUP gives an exotic strong coupling between 
infrared and ultraviolet physics, it resulting in a huge effect at macroscopic scales. To overcome these weakness, in this paper we adopted a new point of view. 
First of all, we have derived, by using heuristic but physically sound physical arguments, Heisenberg uncertainty relations but in a curved spacetime by using tetrad
formalism. Moreover, we used  these so modified Heisenberg uncertainty relations to derive physically motivated STUR in a generic background.
The so obtained STUR are applied to a 
spherically symmetric background. In particular, we have derived expressions for degenerate Fermi momentum and quantum fluctuations. The second part of the paper is devoted to the 
application of these expressions to white dwarf, neutron stars and black holes. As a first result concerning general relativistic effects, we shown that our formulas
confirm that general relativistic effects 
poorly affect the white dwarf physics for equilibrium configuration and critical mass. The situation is more intriguing for more compact objects as neutron stars. There,
general
relativistic effects are very strong. To this purpose, for an extreme degenerate relativistic Fermi neutron gas we derived an interesting approximate
formula, namely (\ref{22c}), relating the critical mass to the compactness
factor $2GM/(c^2R)$. As a result, a minimum critical mass $M_c\simeq 0.59 M_{\odot}$ arises at the Buchdahl limit. Although we used crude approximations to obtain these results,
the fact that our estimation of the Oppenheimer-Volkoff limit is few times smaller \cite{19} than the Newtonian case and that an upper limit for $M$ arises when neutrons are not 
ultra-relativistic, namely equation (\ref{20c}), suggests that effectively our texture of Heisenberg uncertainties in curved spacetimes is physically viable.\\
Concerning quantum gravity effects, contrary to GUP-based results, we found no detectable signatures due to quantum fluctuations for macroscopic 
astrophysical objects. In order to detect quantum effects, we should build self-gravitating systems with mass of the order of the Planck one $M_p$. To this purpose, suppose to have 
a self-gravitating system with mass $\sim M_p$ and density of the order of the one suitable for white dwarf, i.e. $\rho\sim 10^7 kg/m^3$. From the relation
$\rho V\sim M_p$ we deduce that $R\sim 10^{-5}$ meters. As a consequence, it is possible, at least theoretically, to have a self-gravitating system where quantum fluctuations
play an important role. Obviously, at present time, we have not a technology to realize such a laboratory experiment.\\
Finally, we explored possible non commutative effects near a Schwarzschild black hole event horizon. We have obtained the interesting result that at $r=R_s$
spatial STUR disappear and only the time-radius uncertainty remains in spherical coordinates. The possible relation between this phenomenon and the trans Planckian problem 
of Hawking radiation is matter for a next investigation.

\section*{Acknowledgements} 

I'M indebted with Luca Tomassini for many useful discussions and hints concerning the derivation of the STUR and their physical meaning.


\begin{thebibliography}{0}
\bibitem{1}Chandrasekhar S 1935 {\it Mont. Not. of the Royal Astron. Soc.} {\bf 95}(3) 207 
\bibitem{2}Ong Y C 2018 {\it JCAP} {\bf 09}015 
\bibitem{3}Rashidi R 2016 {\it Annals Phys.} {\bf 374} 434 
\bibitem{4} Ong Y C and Yao Y 2018 {\it Phys. Rev. D}{\bf 98} 126018 
\bibitem{5}Wigner E P 1957 {\it Rev. Mod. Phys.} {\bf 29} 255 
\bibitem{6}Mead C A 1964 {\it Phys. Rev.}{\bf 135}B849 
\bibitem{7}Veneziano G 1986 {\it Europhys. Lett.}{\bf 2} 199 
\bibitem{8}Ciafaloni M and Veneziano G 1987 {\it Phys. Lett. B} {\bf 197} 81 
\bibitem{9}Scardigli F 1999 {\it Phys. Lett. B}{\bf 452} 39 
\bibitem{10}Maggiore M 1993 {\it Phys. Lett. B}{\bf 304} 65 
\bibitem{11}Piacitelli G 2010 {\it SIGMA}{\bf 6} 073
\bibitem{12}Kempf A 1994 {\it J. Math. Phys.}{\bf 35}4483 
\bibitem{r1}Scardigli F, Lambiase G and Vagenas E 2017 {\it Phys. Lett B} {\bf 767} 242
\bibitem{r2}Lambiase G and Scardigli F 2018 {\it Phys. Rev. D} {\bf 97} 075003
\bibitem{r3}Buoninfante L, Luciano G G and Petruzziello L 2019 {\it Eur. Phys. J. C} {\bf 79} 663
\bibitem{r4}Luciano G G and Petruzziello L 2019 {\it Eur. Phys. J. C} {\bf 79} 283
\bibitem{13}Lake M J  {\it arXiv:2008.13183}
\bibitem{4a}Doplicher S, Fredenhagen K and  Roberts J E 1995  {\it Comm. Math. Phys.} {\bf 172} 187
\bibitem{14}Bahns D, Doplicher S, Fredenaghen K and  Piacitelli G 2011 {\it Commun. Math. Phys.}{\bf 308} 567 
\bibitem{15}Hossenfelder S 2012 {\it Class. Quant. Grav.}{\bf 29} 115011 
\bibitem{16}Doplicher S, Piacitelli G, Tomassini L and Viaggiu S {\it arXiv:1206.3067}
\bibitem{6a}Tomassini L and Viaggiu S 2014  {\it Class. Quantum Grav.} {\bf 31} 185001
\bibitem{2a}Misner C W, Thorne K S and Wheeler J A 1973 {\it Gravitation, Freeman, San Francisco} 
\bibitem{3a}Fock V A 1926 {\it Z. Phys.} {\bf 39} 226	
\bibitem{5a}Tomassini L and Viaggiu S 2011 {\it Class. Quantum Grav.} {\bf 28} 075001 
\bibitem{7a}Tomassini L and  Viaggiu S {\it in preparation}
\bibitem{8a}Edwards J P 2017 {\it Eur. Phys. J. C} {\bf 77}:320 
\bibitem{9a}Iskauskas A 2015 {\it Phys. Lett B} {\bf 746} 25 
\bibitem{10a}Viaggiu S 2018  {\it Class. Quantum Grav.} {\bf 35} 215011
\bibitem{11a}Viaggiu S 2019  {\it Found. of Phys.} {\bf 49} 1287
\bibitem{12a}Viaggiu S 2019  {\it Physica Scripta.} {\bf 94} 125014
\bibitem{17}Ray A, Maity P and Majumdar P 2019 {\it Eur. Phys. J. C}{\bf 79} 97 
\bibitem{18}Mathew A, Nandy M 2017 {\it Research in Astronomy and Astrophysics} {\bf 17} 061 
\bibitem{19}Ghosh P 2007 {\it Rotation and Accretion Powered Pulsars;} World Scientific Publishing Co.
\end{thebibliography}
\end{document}